\documentclass[nofootinbib,floatfix,superscriptaddress,openany]{iopart}
\usepackage[utf8]{inputenc}
\usepackage{iopams}
\usepackage{amsfonts}
\usepackage{amssymb}
\usepackage{color}
\usepackage{showlabels}
\usepackage{graphicx}
\usepackage{hyperref}

\begin{document}

\title[Evolution of initially contracting Bianchi Class A models]{Evolution
of initially contracting Bianchi Class A models in the presence of an
ultra-stiff anisotropic pressure fluid}
\author{John D. Barrow and Chandrima Ganguly}

\begin{abstract}
We study the behaviour of Bianchi class A universes containing an
ultra-stiff isotropic ghost field and a fluid with anisotropic pressures
which is also ultra-stiff on the average. This allows us to investigate
whether cyclic universe scenarios, like the ekpyrotic model, do indeed lead
to isotropisation on approach to a singularity (or bounce) in the presence
of dominant ultra-stiff pressure anisotropies. We specialise to consider the
closed Bianchi type IX universe and show that when the anisotropic pressures
are stiffer on average than any isotropic ultra-stiff fluid then, if they
dominate on approach to the singularity, it will be anisotropic. We include
an isotropic ultra-stiff ghost fluid with negative energy density in order
to create a cosmological bounce at finite volume in the absence of the
anisotropic fluid. When the dominant anisotropic fluid is present it leads
to an anisotropic cosmological singularity rather than an isotropic bounce.
The inclusion of anisotropic stresses generated by collisionless particles
in an anisotropically expanding universe is therefore essential for a full
analysis of the consequences of a cosmological bounce or singularity in
cyclic universes.
\end{abstract}

\ead{J.D.Barrow@damtp.cam.ac.uk}

\ead{C.Ganguly@damtp.cam.ac.uk} 
\bigskip
\address{DAMTP, Centre for Mathematical Sciences, University of Cambridge, Cambridge
CB3 0WA, United Kingdom}

\maketitle

\section{Introduction}

The standard model of cosmology has been subjected to detailed scrutiny by
recent WMAP and Planck mission data. It predicts an almost isotropic,
homogeneous and flat expanding universe, an approximately scale-invariant
inhomogeneity spectrum with some level of statistical non-gaussianity, and
observational parameters linked by an underlying inflationary model for its
very early history. This inflationary model requires an early period of
accelerated expansion to account for the horizon and flatness problems, and
to generate density perturbations that seed the formation of galaxies.
Despite the success of the inflationary paradigm, it has little to say about
the initial state of the universe and if, or how, a big bang singularity
prior to inflation might be avoided or mitigated.

Problems such as these have prompted the search for alternatives to
inflation or natural initial conditions that lead to inflation. The
philosophy behind these searches is that although inflation is a very
successful theory, it is important to search for alternative theories which
can provide the similar predictions as inflation, yet which might be
distinguished by some decisive observations. One of the oldest alternatives
is that of a non-singular bounce.

The existence of a non-singular bounce which facilitates the transition from
an initially contracting universe to an expanding one was first hypothesized
in general-relativistic cosmology by Tolman and Lema\^{\i}tre \cite%
{Lemaitre1997,Tolman1931} and was updated to include more general aspects of
general relativistic cosmology and the presence of a cosmological constant
by Barrow and Dabrowski \cite{BarrDab}, \cite{review}. It also regained
popularity in the context of pre-big-bang scenarios \cite{Gasperini2002},
which although not successful, led to developments in theories which could
possess a non-singular bounce. These are usually produced by the addition to
standard cosmology of an effective field which violates the null energy
condition (NEC). For example, cosmologies with ghost condensates or Galilean
genesis take this approach \cite{Creminelli2010}. This has also been used
effectively in quantum editions of cosmology, especially in loop quantum
cosmology \cite{Singh2006}, in theories involving canonical quantization of
gravity \cite{CASADIO2000,AcaciodeBarros1998}, or classical theories with
varying constants \cite{vary1}, \cite{vary2} and ghost fields \cite{Btsagas}.

One of the first questions to ask when considering alternatives (or
additions) to standard inflation is whether it solves the problems that
inflation claims to solve. For example, one can ask whether the present-day
isotropy and homogeneity of the universe can be achieved through a cosmology
which underwent contraction and bounce at some time (or times) in the past.
It has been claimed that models implementing a phase of ekpyrosis, or a
phase of scalar field-driven fast contraction can indeed solve this problem 
\cite{Lehners2008}, \cite{Khoury2001}, \cite{Buchbinder2007}. In effect,
this model claims to solve the anisotropy problem by introducing a scalar
field with negative potential energy, which behaves as an ideal fluid with
ultra-stiff equation of state $p\gg\rho $. Its isotropic density therefore
grows faster than the anisotropies in a contracting universe because the
latter diverge no faster than an effective $p=\rho $ fluid. However, this
simple analysis assumes that the matter pressure distribution is isotropic.
A full analysis needs to include the effects of matter sources with
anisotropic pressure distributions on approach to the singularity. Since the
isotropic pressure is assumed to exceed the energy density, it should be
permitted for the average pressure to exceed the energy density of the
anisotropic fluid as well. The need to include anisotropic pressures on
approach to the singularity is important because interactions all become collisionless at a higher temperature in the case of anisotropic expansion, than in the isotropic case. Their 
interaction rates can be written as $\Gamma =\sigma n v \sim g\alpha ^{2}T,$, where
$\sigma$ is the interaction cross section, $n$ is the number density of particles, $v$ is the
average velocity of the particles, $\alpha$ is the generalised structure constant associated with any interaction mediated by some gauge boson, $T$ is the
temperature of the universe and $g$ is the effective number of relativistic degrees of freedom of particles at the temperature $T$ . This interaction rate will be slower than the cosmological expansion rate, $H\simeq g^{1/2}T^{2}m_{p}^{-1}$, whenever $%
T>g^{1/2}\alpha ^{2}m_{p}\sim 10^{16}GeV$ in simple unified models. In the preceding line $m_p$ is the Planck mass. If the
expansion is anisotropic down to a temperature $T_{F}$, the expansion rate
is faster, by a factor of $T/T_{F}$. As $\Gamma/H =(T_{F}/T^2)g^{1/2}\alpha ^{2}m_p$ at $T>T_F$, collisional equilibrium is even harder to maintain when $T>T_{F}$. Graviton production near $T=m_{p}$ also produces a population of collisionless particles whose free streaming will produce significant
anisotropic pressures if the expansion dynamics are anisotropic \cite{starob}%
, \cite{skew}. 

In this paper, we investigate the effects of anisotropic pressures in the
Bianchi Class A homogeneous, anisotropic cosmologies, generalising the study
of these effects in the simple Bianchi type I cosmological model by Barrow
and Yamamoto \cite{BYam}. We carry out a generalized phase-plane analysis
for all the cosmologies of this type but then focus on the closed Bianchi
type IX cosmologies and carry out numerical calculations to study their
behaviour near any initial singularity or expansion minimum when this kind
of anisotropic matter content is present in addition to an ultra-stiff
isotropic fluid. We will show that in these most general homogeneous and
anisotropic cosmologies it is essential to include the effects of
anisotropic pressures as well as shear anisotropy. When the anisotropic
pressures are stiffer on average than the isotropic pressures then they
determine the nature of any singularity (or bounce) and it will be dominated
by anisotropy, contrary to the situation expected in the standard ekpyrotic
picture which ignores anisotropic pressures.

This paper is organised as follows. We begin by presenting the generalised
Einstein field equations in an expansion-normalised dynamical system for the
non-tilted Bianchi Class A models containing isotropic ultra-stiff ($p>\rho $%
) matter content as well as a second ultra-stiff matter source with positive
density and anisotropic pressures. We first perform a stability analysis on
this system for an initially contracting universe to see if a phase of
ekpyrosis is really successful in suppressing the anisotropies in the
presence of a dominant anisotropic pressure fluid. We also seek solutions to
these equations in the limit of small anisotropy and give a new Bianchi I
exact solution. In the next section, we study explicitly the evolution of a
contracting, anisotropic but spatially homogeneous universe near the initial
singularity in the presence of the matter content prescribed, then
specializing to the Bianchi type IX universe. We then show the results of
our numerical calculations in this universe and compare our results to the
results of the stability analysis of the previous section. In the last
section we draw our conclusions.

\section{Ekpyrotic models}

In this section, we first review the simplest ekpyrotic models and the way
they suppress the dominant growth of anisotropies in a contracting universe
as it approaches the singularity. The ekpyrotic models \cite%
{Khoury2001,Donagi2001} were originally based on a $5$-dimensional
braneworld scenario, where the fifth dimension ends at two boundary branes,
one of them being our universe. The branes could interact with each other
only gravitationally and are attracted by inter-brane tension during the
phase of ekpyrosis. Thus, the universe underwent a phase of slow contraction
before the collision and re-expansion of the branes, an event which was
identified with the hot big bang. The branes were not completely uniform.
Quantum fluctuations cause their collision to occur at different times in
different places. Thus some parts of the universe end up hotter than others,
giving rise to density and temperature fluctuations. This model has been
criticized due to fine tuning problems \cite{Kallosh2001}, problems
regarding the contracting phase seeming to end in a singularity \cite%
{Kallosh2001a}, and also because of its failure to produce a scale-invariant
spectrum of density fluctuations \cite{Martin2003}. To circumvent such
problems, some modifications were proposed in terms of a cyclic model \cite%
{Steinhardt2002a,Steinhardt2002,Khoury2002}, with alternating phases of
contraction (when the branes approach each other) and expansion (when the
branes are pulled apart and the universe enters a phase of dark-energy
domination) occurring simultaneously in a cycle. New versions still attract
debate about fundamental issues \cite{bars, kallosh1}. The turnaround from
contraction to expansion was hypothesized to occur in the form of a
non-singular bounce facilitated by a ghost condensation mechanism \cite%
{Buchbinder2007}. Furthermore, it was seen that a scale-invariant density
fluctuation power-spectrum could be generated in the new ekpyrotic scenario
if one considered two-field ekpyrosis \cite{Buchbinder2007}. These
possibilities have sustained interest in the ekpyrotic scenario as an
alternative to inflation for the origin of structure in the universe. If
primordial gravitational waves were reliably detected \cite{bic} then their
amplitude could provide a decisive test between the two alternatives (and
others \cite{brand}).

\subsection{Effects on expansion anisotropy}

As mentioned above, this model also claims to solve the problem of growing
shear by incorporating the ekpyrotic phase \cite{Lehners2008,Buchbinder2007}%
. This ekpyrotic phase has also been used in other cosmological bouncing
models as a way to deal with the problem of growing anisotropies in a
contracting universe \cite{Cai2013a}, and so merits closer investigation.
For simplicity, we shall focus on the single-field ekpyrotic model and first
describe the effects of ekpyrosis in a Bianchi I universe with the ekpyrotic
field and an ultra-stiff energy source with anisotropic pressure. The
ekpyrotic field is a scalar field, $\phi $ rolling down rapidly on a steep
negative potential. This can be viewed as driving the contraction of the
universe. To see how it might suppress the anisotropies, we write down its
effective equation of state \cite{Lehners2008}.

\begin{equation}
p=(\gamma -1)\rho ,\;\;\;\;\;\;\;\;\;\;\;\gamma \gg2.
\end{equation}%
The anisotropy energy density scales as $1/l^{6}$, and behaves like a source
with $\gamma =2$, $l$ being the time dependent mean scale factor of the
universe. Thus, the ekpyrotic phase simply introduces a source which scales
with scale factor faster than the energy density in the anisotropy because $%
\gamma >2$, see \cite{coley}. As the universe contracts, this term dominates
over the anisotropy in the Friedmann equation, apparently solving the
problem of isotropising the universe before it enters the hot big bang phase
-- or at least preventing the new expanding phase beginning with highly
anisotropic dynamics. This should also result in significant dissipation and
particle production which would reduce anisotropy and generate entropy \cite%
{starob}, \cite{bmatz}. We ignore these complicated effects here.

The simplest form of an anisotropic but spatially homogeneous universe is
the Bianchi I (or Kasner) universe \cite{kas}, \cite{taub}. The metric is 
\begin{equation}
ds^{2}=-dt^{2}+a^{2}(t)dx^{2}+b^{2}(t)dy^{2}+c^{2}(t)dz^{2},  \label{one}
\end{equation}%
The Einstein equations for this model gives \cite{WEllis} 
\begin{equation}
3H^{2}=\sigma ^2+\rho _{matter},
\label{friedmann_kasner}
\end{equation}%
\begin{equation}
\dot{\sigma}_{\alpha \beta }+3H\sigma _{\alpha \beta }=\mu \mathcal{P}%
_{\alpha \beta }  \label{shear_evolution_kasner}
\end{equation}%
where $\sigma _{\alpha \beta }$ is the shear tensor which follows the relation
\begin{equation}
\sigma^2=\frac{1}{2}\sigma ^{\alpha \beta}\sigma _{\alpha \beta}
\end{equation}
and  $\mu $ is the
anisotropic pressure fluid density and the definition of $\mathcal{P}%
_{\alpha \beta }$ is shown in \Eref{Palphabeta}. Also $\rho_{matter}$ refers
to the total energy density of the matter components of the system, i.e., the matter
with isotropic pressures as well as the matter with anisotropic pressure.
If we have only a fluid with isotropic pressure, then the right-hand side of %
\Eref{shear_evolution_kasner} vanishes and we can write the shear energy
density, $\sigma ^2$ in the Friedmann
constraint, \Eref{friedmann_kasner} as $\Sigma ^{2}/l^{6}$, where $\Sigma
^{2}$ is constant. Hence, an ekpyrotic field with equation of state, $%
p_{\phi }\gg\rho _{\phi }$ would dominate over the anisotropy when $%
l\rightarrow 0$ and the singularity is approached. We can give a new exact
Bianchi type I solution of Einstein's equations in a form which illustrates
this in the particular representative case with $\mathcal{P}_{\alpha \beta
}=0$ and $p=3\rho $ where the metric is exactly integrable (\Eref{one}):

\begin{equation}
a(t)=\left( (t^{2}+C_{2}t)^{1/2}(\sqrt{t}+\sqrt{(C_{2}+t)}%
)^{2(3q_{1}-1)}\right) ^{1/3}  \label{k1}
\end{equation}%
\begin{equation}
b(t)=\left( (t^{2}+C_{2}t)^{1/2}(\sqrt{t}+\sqrt{(C_{2}+t)}%
)^{2(3q_{2}-1)}\right) ^{1/3}  \label{k2}
\end{equation}%
\begin{equation}
c(t)=\left( (t^{2}+C_{2}t)^{1/2}(\sqrt{t}+\sqrt{(C_{2}+t)}%
)^{2(3q_{3}-1)}\right) ^{1/3}  \label{k3}
\end{equation}

\begin{equation}
\sum_{i}q_{i}=1=\sum_{i}q_{i}^{2}  \label{k4}
\end{equation}%
Thus, we see that at early times this solution tends to the flat Friedmann
solution for $p=3\rho $ 'matter': $a\sim t^{1/6}$, $b\sim t^{1/6}$and $c\sim
t^{1/6}$ as $t\rightarrow 0$), and at late times approaches the Kasner
solution $a\sim t^{q_{1}}$, $b\sim t^{q_{2}}$ and $c\sim t^{q_{3}},$with
condition \Eref{k4} as $t\rightarrow \infty;$ fuller details can be found in
the Appendix. Thus, this solution provides a simple description of the
transition from an isotropic initial state to a Kasner-like anisotropic
future. This is the opposite trend to the evolution of a $p<\rho $
perfect-fluid model.

However, if we relax the assumption of having energy sources with only
isotropic pressure, we can no longer write down the form of the anisotropy
energy density in \Eref{friedmann_kasner} the simple form, $\Sigma
^{2}/l^{6},$ since the right-hand side of \Eref{shear_evolution_kasner} no
longer vanishes. In fact, the anisotropy may diverge faster than the
ekpyrotic fluid in powers of $l^{-1}$ in any particular direction as $%
t\rightarrow 0$, depending on the pressure component of the matter source in
that direction. Hence, we can no longer be sure that adding a matter
component with $w\gg1$ solves the problem of isotropising the universe on
approach to the singularity. This will be investigated in more detail and
for more general forms of anisotropic spatially homogeneous universes in 
\Sref
{sec:stability_analysis}.

\section{Bianchi Class A models of types I-VIII}

In this section, we investigate the assumption that an ultra-stiff energy
source suppresses the anisotropies near a singularity in an initially
contracting universe. We do this for the Bianchi Class A models \cite%
{EllisMac}, which generalise the Bianchi type I models because they allow
the presence of anisotropic spatial curvature. However, now we add an
ultra-stiff anisotropic pressure source comoving with the isotropic fluid
source to see if ekpyrosis still manages to suppress the anisotropies. The
investigation of whether the anisotropies are suppressed by the ekpyrotic
phase has been done in the case of an empty anisotropic spatially
homogeneous geometry, the Kasner universe \cite{Erickson2004}, but without
the anisotropic pressure fluid. Studies regarding the inclusion of an
anisotropic fluid in the Bianchi universes have also been made in \cite%
{Calogero2011}. However, here we follow the approach similar to the one used
in \cite{BYam,Lidsey2005} and present a more general analysis for all the
Bianchi models included in Class A with the aim of finding the conditions
under which the Friedmann-Lema\^{\i}tre (FL) fixed point is an attractor for
a contracting universe on approach to the collapse.

The ultra-stiff isotropic matter considered in this section is a
null-energy-condition-violating fluid with negative energy density. This
negative energy density is introduced to induce a bounce at early times
instead of a singularity. This is because many bouncing cosmologies consider
fields that effectively behave as a ghost field to facilitate the bounce and
which also behave as a stiff or ultra-stiff matter source \cite%
{Btsagas,Buchbinder2007,CASADIO2000}.

We can write the energy-momentum tensor as follows: 
\begin{equation}
T_{ab}^{total}=T_{ab}^{I}+T_{ab}^{A},
\end{equation}%
where the superscripts, $I$ and $A$ denote "isotropic" and "anisotropic"
respectively. The anisotropic fluid energy-momentum tensor can be written
explicitly as 
\begin{equation}
T_{ab}^{A}=\mu \{u_{a}u_{b}+(\gamma _{\star }-1)(g_{ab}+u_a u_b)+\mathcal{P}_{ab}\}.
\label{Palphabeta}
\end{equation}%
In the rest of this work, the isotropic and anisotropic fluid energy
densities will be referred to as $\rho $ and $\mu $ respectively and the
isotropic fluid will have equation of state $p=(\gamma -1)\rho $ while the
anisotropic pressure tensor $\mathcal{P}_{ab}$ has diagonal elements $%
(\gamma _{i}-\gamma_{\star})$, for all $i=1,2,3$, respectively, with average value $%
\gamma _{\star }=(\gamma _{1}+\gamma _{2}+\gamma _{3})/3$. The isotropic
fluid energy density, $\rho $, is that of a ghost field and so $\rho <0$.

We can write the Einstein equations for this class of cosmological models
with the specified matter content as follows \cite{WEllis}: 
\begin{equation}
\dot{H}=-H^{2}-\frac{2}{3}\sigma ^{2}-\frac{1}{6}(\rho +3p)-\frac{1}{6}(\mu
+p_{1}+p_{2}+p_{3}),
\label{efe1}
\end{equation}%
where the anisotropic pressures are defined to be $p_i =(\gamma _i -1)\mu$ for $i=1,2,3$.
\begin{equation}
\dot{\sigma}_{\alpha \beta }=-3H\sigma _{\alpha \beta }+2\epsilon _{(\alpha
}^{\mu \nu }\sigma _{\beta )\mu }\Omega _{\nu }-^{(3)}S_{\alpha \beta }+
\mathcal{P}_{\alpha \beta } \, \mu ,  \label{shear_1}
\end{equation}%
\begin{equation}
\dot{n}_{\alpha \beta }=-Hn_{\alpha \beta }+2\sigma _{(\alpha }^{\mu
}n_{\beta )\mu }+2\epsilon _{(\alpha }^{\mu \nu }n_{\beta )\mu }\Omega _{\nu
}
\label{efe2}
\end{equation}%
\begin{equation}
\dot{\rho}=-3\gamma H\rho ,  \label{continuity_equation_1}
\end{equation}%
\begin{equation}
\dot{\mu}=-3\gamma _{\star }H\mu -\sigma _{\alpha \beta }\mathcal{P}^{\beta
\alpha }\mu .  \label{continuity_anisotropic_1}
\end{equation}%
The Friedmann constraint is given by 
\begin{equation}
H^{2}=\frac{\rho }{3}+\frac{\mu }{3}+\frac{\sigma ^{2}}{3}-\frac{^{(3)}R}{6}.
\label{friedman_constraint_1}
\end{equation}

We introduce, 
\begin{eqnarray*}
\sigma _{+}& \equiv \frac{1}{2}(\sigma _{22}+\sigma _{33}), \\
\sigma _{-}& \equiv \frac{1}{2\sqrt{3}}(\sigma _{22}-\sigma _{33}).
\end{eqnarray*}

Diagonalising the stress tensor we find that all other components of the
stress evolution equation are not dynamical. Similarly, the trace-free
spatial Ricci tensor $^{(3)}S_{\alpha \beta }$ and the constant tensor $%
\mathcal{P}_{\alpha \beta }$ are diagonal and their components $\mathcal{P}%
_{+}$ and $\mathcal{P}_{-}$ and $S_{+}$ and $S_{-}$ are given by
analogous expressions. Explicitly, the expansion-normalized combinations of
the dynamical components of the Ricci tensor are given by. 
\begin{eqnarray}
\mathcal{S}_{+} &=&\frac{1}{6H^{2}}[(n_{2}-n_{3})^{2}-n_{1}(2n_{1}-n_{2}-n_{3})],
\\
\mathcal{S}_{-} &=&\frac{1}{2\sqrt{3}H^{2}}(n_{3}-n_{2})(n_{1}-n_{2}-n_{3}).
\end{eqnarray}%
where the $n_i$ are the principal components of the structure tensor $n_{\alpha \beta}$.
The scalar curvature is given by, 
\begin{equation}
^{(3)}R=-\frac{1}{2}%
[n_{1}^{2}+n_{2}^{2}+n_{3}^{2}-2(n_{1}n_{2}+n_{2}n_{3}+n_{3}n_{1})].
\end{equation}%
We want to set up the phase space so that we can study the evolution of the
quantities with respect to the expansion of the universe, i.e.,\ with respect
to $H\equiv \dot{l}/l$ where $l(t)$ is a generalised mean scale factor. We
begin by introducing expansion-normalised variables as follows, 
\begin{equation}
\fl\Sigma_{\pm}\equiv \frac{\sigma _{\pm}}{H},\quad N_i\equiv \frac{n_i}{H},\quad \Omega \equiv \frac{-\rho }{3H^{2}},\quad Z\equiv \frac{\mu }{3H^{2}},\quad
\Sigma ^{2}\equiv \frac{\sigma ^{2}}{3H^{2}},\quad K\equiv -\frac{^{(3)}R}{%
6H^{2}}
\end{equation}%
\footnote{%
The minus signs in the definition of $\Omega $ and $K$ ensure that they are
positive when $\rho <0$ and $^{(3)}R<0$ in our system} The Bianchi Class A
universe will now be determined completely if we solve the Einstein's field
equations in these new variables for the state vector $\{H,\Sigma
_{+},\Sigma _{-},N_{1},N_{2},N_{3},\Omega ,Z\}$. We find that the Friedmann
constraint\eref{friedman_constraint_1} becomes, 
\begin{equation}
-\Omega +Z+\Sigma ^{2}+K=1
\end{equation}%
where $\Sigma ^2 =\Sigma _- ^2 +\Sigma _+^2$.
In terms of the expansion-normalised variables, the Einstein field equations
become:

\begin{eqnarray}
\fl \Sigma_{\pm}'&=-(2-q)\Sigma_{\pm}-\mathcal{S}_{\pm}+3P_{\pm}Z\\
\fl N_1'&=(q-4\Sigma_+)N_1\\
\fl N_2'&=(q+2\Sigma_++2\sqrt{3}\Sigma_-)N_2\\
\fl N_3'&=(q+2\Sigma_+-2\sqrt{3}\Sigma_-)N_3\\
\fl \Omega'&=[3(\gamma_{\star}-\gamma)-(3\gamma_{\star}-2)K+3(2-\gamma _{\star})\Sigma^2+3(\gamma_{\star}-\gamma)\Omega]\Omega\\
\fl Z'&=[3(2-\gamma _{\star})\Sigma^2-3(\gamma-\gamma_{\star})\Omega-(3\gamma _{\star}-2)
K-6(P_+\Sigma_++P_-\Sigma_-)]Z
\end{eqnarray}%
Here, all time derivatives $^{\prime }$ are taken with respect to a new time
coordinate $\tau $ which is defined by the relation, 
\begin{equation}
\frac{dt}{d\tau }=\frac{1}{H}=\frac{l}{\dot{l}},
\end{equation}%
The deceleration parameter, $q=-\ddot{l}l/\dot{l}^{2}$ is given by 
\begin{equation}
\dot{H}=-(1+q)H^{2}
\end{equation}%
In the expansion-normalized variables, using the Friedmann constraint, we
find that $q$ is given by 
\begin{equation}
q=2\Sigma ^{2}-\frac{1}{2}(3\gamma -2)\Omega +\frac{1}{2}(3\gamma _{\star
}-2)Z.
\end{equation}%
One degree of freedom has been removed in the system by using the Friedmann constraint to substitute $Z$ in the evolution equation for $\Omega$ to obtain a $6$ dimensional system rather than the original $7$ dimensional system, as no new information can be obtained by considering the linearised version of the evolution equation for $Z$ around any of the fixed points.
\begin{table*}[t]
\caption{Equilibrium points of the Bianchi I-VIII dynamical systems}$%
\centering
\begin{array}{|c|ccccccc|}
\hline
\, & \Sigma_+ & \Sigma_- & N_1 & N_2 & N_3 & \Omega & Z \\ \hline
FL & 0 & 0 & 0 & 0 & 0 & -1 & 0 \\ 
P_1^+(II) & \Sigma^{\!^{(1)}}_{\!_{P_{II}}} & 0 & n^{\!^{(1)}}_{\!_{P_{II}}}
& 0 & 0 & \Omega^{\!^{(1)}}_{\!_{P_{II}}} & 0 \\ 
P_1^+(VI_0) & \Sigma^{\!^{(1)}}_{\!_{P_{VI}}} & 0 & 0 & n^{\!^{(1)}}_{%
\!_{P_{VI}}} & -n^{\!^{(1)}}_{\!_{P_{VI}}} & \Omega^{\!^{(1)}}_{\!_{P_{VI}}}
& 0 \\ 
Kasner & \Sigma^{\!^2}_{\!_{K-}} & \Sigma^{\!^2}_{\!_{K+}} & 0 & 0 & 0 & 0 & 
0 \\ 
\mathcal{A}_1 & \frac{2}{2-\gamma_{\star}}P_+ & \frac{2}{2-\gamma_{\star}}P_-
& 0 & 0 & 0 & 0 & (1-\Sigma^2_{\!_{(\mathcal{A}_1)}}) \\ 
\mathcal{A}_2 &\Sigma _+^{\mathcal{A}_2} & \Sigma_-^{\mathcal{A}_2} & 0 & 0 & 0 & \Omega ^{\mathcal{A}_2} &Z^{\mathcal{A}_2} \\ \hline
\end{array}
$%
\label{fixed_points}
\end{table*}

\subsection{\label{sec:stability_analysis}Stability analysis}

We consider the evolution equations in the expansion-normalised variables
and perform a phase-plane analysis for them. We first identify the
equilibrium points. In \Tref{fixed_points}, the explicit forms of the
quantities referred to in the Table are given in the Appendix. On
examination of these expressions, we find that for cases of ultra-stiff
matter, as well as for cases when the anisotropic fluid is stiffer than the
isotropic fluid, all these points become unphysical except for the $FL$, $%
Kasner$, $\mathcal{A}_{1}$ and $\mathcal{A}_{2}$ points.

We are interested in understanding the behaviour of this system of equations
with respect to the FL fixed point ($FL$) in the asymptotic past. Thus, we
linearise this system of equations around this fixed point to obtain the following equations: 
\begin{eqnarray}
\Sigma _{\pm}^{\prime } &=&-\frac{3}{2}(\gamma -2)\Sigma _{\pm}+3\mathcal{P}%
_{\pm}Z, \\
N_{i}^{\prime } &=&\frac{1}{2}(3\gamma -2)N_{i}, \;\;\;\forall i=1,2,3 \\
\Omega ^{\prime } &=&3(\gamma -\gamma _{\star })\Omega, \\
\end{eqnarray}%
Thus the eigenvalues are $(3\gamma -2)/2$ of multiplicity $3$, $(3(2-\gamma)/2$ of multiplicity $2$ and in the reduced system, the remaining eigenvalue is $3(\gamma -\gamma _{\star})$. Using the condition $\gamma _{\star}>\gamma >2$, we find that the FL point is future stable and future unstable in certain directions. Thus a further analysis needs to be undertaken to determine the behaviour in the presence of an ultra stiff anisotropic pressure fluid.

\section{Bianchi IX universe with isotropic ghost field and fluid with
anisotropic pressures}

\label{sec:bianchi_ix}

We now consider the specific example of an anisotropic but spatially
homogeneous closed universe of Bianchi type IX. This is the most interesting
case because it contains the closed isotropic FL universe as a special case.
It also displays the most general chaotic dynamics on approach to the
singularity \cite{Misner1969}, \cite{Belinskii1970} in the absence of stiff
or super-stiff matter fields. It was assumed in this work that the matter
and radiation sources could be neglected near the singularity and the vacuum
dynamics is asymptotically approached for any perfect fluid source with $%
0\leq p<\rho $. As is well known, the chaotic type IX evolution is
well-approximated by an infinite succession of Kasner epochs, which occur in
any finite open interval of proper time around $t=0$. At any instant two of
the scale factors oscillate with approximate Kasner initial conditions at
the beginning of each epoch while the third decreases monotonically with
time as $t\rightarrow 0$ \cite{LLif}. The sequence of oscillatory Kasner
configurations appears to be chaotic in nature and the discrete dynamics can
be solved exactly to find the smooth invariant measure \cite{BChernoff,
barIX}. It is a non-separable measure of the sort that characterises a
double-sided continued fraction map. However, the inclusion of a stiff
matter fluid with equation of state $p=\rho $ \cite{Belinskii1972}, \cite%
{matz} results in an inevitable termination of the chaotic oscillations on
approach to the singularity, after which all three scale factors evolve
monotonically (but not in general isotropically) to zero as $t\rightarrow 0$
because the Kasner solution for $p=\rho $ matter permits all the Kasner
indices to be simultaneously positive and the initial state is
quasi-istropic \cite{JBquiet}. The chaotic oscillatory sequence towards the
singularity ends: no further oscillations occur.

Thus, the inclusion of a stiff matter fluid in the Bianchi IX system
ultimately suppresses the chaotic behaviour of the scale factors near the
singularity. In the Misner's Hamiltonian picture this corresponds to the
universe point eventually having too low a perpendicular velocity component
relative to the potential wall it is approaching as the walls expand on
approach to the singularity. It never reaches the wall and remains moving as
if there are no potential walls (ie as in a Bianchi type I universe). No
further transpositions of Kasner behaviours occur. \emph{\ }All the known
ways in which chaotic behaviour can be avoided in type IX universes exploit
this feature directly or indirectly and are linked to the dimension of space
in an interesting way \cite{dem}. Clearly, in the ekpyrotic scenario \cite%
{Erickson2004}, a phase of ekpyrotic evolution, which is equivalent to
domination by an ultra-stiff fluid with $p>\rho ,$ will have a more
pronounced effect of suppressing the anisotropy energy domination and
driving the dynamics towards isotropy. We investigate if this conclusion is
sustained in the presence of anisotropic pressures.

The type IX universe is also interesting because it reduces to the closed FL
universe in the isotropic limit, and this has been shown to possess very
simple cyclic behaviour in the presence of ghost stiff matter content ($%
p=\mu <0$) and radiation \cite{Btsagas}. This model therefore seems to be a
suitable candidate to test the results of our stability analysis of the
previous section and also to learn more about the explicit behaviour of the
scale factors, their Hubble rates, and the shear anisotropy tensor.

The matter considered in the following analysis is, as before, an
ultra-stiff ghost field plus a stiffer anisotropic pressure field. The ghost
field is included because, if it dominates at small times, it will create a
bounce at a non-zero expansion volume minimum. The dynamics will be driven
towards isotropy if a bounce occurs. By contrast, if the ultra-stiff
anisotropic pressure field dominates over the isotropic ghost field then it
should drive the dynamics towards an anisotropic Weyl curvature singularity.

\subsection{Field equations}

In this section we analyse a diagonal Bianchi type IX universe containing an
isotropic ultra-stiff ghost field (with negative density) and another fluid
with positive density and anisotropic pressures. We will test the
possibility that on approach to a singularity the ultra-stiff ghost field
will dominate over the anisotropic pressures and so cause the universe to
isotropise and bounce at a finite expansion minimum. However, if the average
anisotropic pressure becomes larger than that of the ghost field then we
expect the singularity to be restored because the anisotropic pressures will
dominate the dynamics at the singularity. The ghost field is\ included here
simply as a device to bring about a simple bounce at finite radius; however,
some editions of the ekpyrotic scenario do include an effective ghost by
allowing the sign of the gravitational coupling to change because it is
determined by a time-dependent scalar field \cite{kallosh1, bars}.

In the following, we consider a Bianchi IX universe with scale factors $%
a(t),b(t)$ and $c(t)$, containing an isotropic, ultra-stiff ghost field (negative energy
density, $p>\rho $) as well as an
anisotropic ultra-stiff field, with average stiffness exceeding that of the
isotropic ghost field. The energy density of the isotropic ghost field is
given by $\rho $ and it has pressure $p$ with equation of state, 
\begin{equation}
p=(\gamma -1)\rho ,
\end{equation}%
and the ultra-stiff condition requires $\gamma >2$. The energy density of
the anisotropic pressure ``fluid" is denoted by $\mu ,$ as before. The
equation of state in the $i$th direction, where $i=1,2,3$ and denotes the $3$
spatial directions, is given by, 
\begin{equation}
p_{i}=(\gamma _{i}-1)\mu ,
\end{equation}%
and the ultra-stiff condition requires that some of the $\gamma _{i},$ or
their mean value, exceed $2$. As is evident from the above equations, the $%
p_{i}$ are the anisotropic pressures in the three orthogonal expansion
directions. In general, none of the $p_{i}$'s will be equal. The field
equations for such a type IX universe are: 
\begin{eqnarray*}
\fl \frac{\ddot{a}}{a}+\frac{\ddot{b}}{b}+\frac{\dot{a}\dot{b}}{ab}+\frac{%
a^{2}}{4b^{2}c^{2}}+\frac{b^{2}}{4a^{2}c^{2}}-\frac{3c^{2}}{4a^{2}b^{2}}+%
\frac{1}{2a^{2}} +\frac{1}{2b^{2}}-\frac{1}{2c^{2}}& =-(p+p_{3}) \\
=-(\gamma -1)\rho -(\gamma _{3}-1)\mu , \\
\fl \frac{\ddot{b}}{b}+\frac{\ddot{c}}{c}+\frac{\dot{b}\dot{c}}{bc}+\frac{%
b^{2}}{4a^{2}c^{2}}+\frac{c^{2}}{4a^{2}b^{2}}-\frac{3a^{2}}{4b^{2}c^{2}}+%
\frac{1}{2b^{2}} +\frac{1}{2c^{2}}-\frac{1}{2a^{2}}& =-(p+p_{1}) \\
=-(\gamma -1)\rho -(\gamma _{1}-1)\mu , \\
\fl \frac{\ddot{c}}{c}+\frac{\ddot{a}}{a}+\frac{\dot{c}\dot{a}}{ca}+\frac{%
a^{2}}{4b^{2}c^{2}}+\frac{c^{2}}{4a^{2}b^{2}}-\frac{3b^{2}}{4a^{2}c^{2}}+%
\frac{1}{2a^{2}} +\frac{1}{2c^{2}}-\frac{1}{2b^{2}}&=-(p+p_{2}) \\
=-(\gamma -1)\rho -(\gamma _{2}-1)\mu
\end{eqnarray*}

\begin{equation}
\fl \frac{\dot{a}\dot{b}}{ab}+\frac{\dot{b}\dot{c}}{bc}+\frac{\dot{c}\dot{a}%
}{ca}+\frac{1}{2a^{2}}+\frac{1}{2b^{2}}+\frac{1}{2c^{2}}-\frac{a^{2}}{%
4b^{2}c^{2}}-\frac{b^{2}}{4a^{2}c^{2}}-\frac{c^{2}}{4a^{2}b^{2}}=\rho +\mu .
\label{friedmann_constraint_bianchi_ix}
\end{equation}

We note that these equations can be solved exactly in the special case of
the axisymmetric Bianchi IX universe, with $b=c$ and also $%
p_{1}=p_{2}=p_{3}=\rho ,$ and for other equivalent cyclic permutations. The
solutions have the form, 
\begin{eqnarray*}
a(\tau )^2 = A\mathrm{sech}(A\tau ) \\
b(\tau )^2=\frac{B^2}{4A}\mathrm{cosh}(A\tau)\mathrm{sech}^2\left(\frac{B}{2}\tau\right)
\end{eqnarray*}
subject to the constraints,
\begin{eqnarray*}
\;\;\;\;\;\;\rho=\frac{M^2}{4 a^2 b^2}\\
A^2+M^2=B^2
\end{eqnarray*}%
where $dt=ab^2d\tau $ \cite{JBquiet}.

For the general ekpyrotic case with $p_{1}\neq p_{2}$, an exact solution is
unobtainable. Thus, we resort to finding a numerical solution for the full
Bianchi IX evolution, including both ultra-stiff isotropic and anisotropic
pressures. To facilitate the numerical integration, the scale factors in the
three directions are rewritten in terms of their logarithms as \cite%
{Belinskii1970} 
\begin{equation}  \label{expscale}
a(t)\equiv \mathrm{e}^{\alpha (t)},b(t)\equiv \mathrm{e}^{\beta
(t)},c(t)\equiv \mathrm{e}^{\delta (t)}.
\end{equation}

The field equations can be rewritten as a first-order system by an
appropriate choice of variables (see Appendix for details). Three new
quantities are introduced to achieve this: 
\begin{eqnarray}
x& \equiv \alpha ^{\prime }(t)-\beta ^{\prime }(t), \\
y& \equiv \alpha ^{\prime }(t)-\delta ^{\prime }(t), \\
H& =\frac{1}{3}\left( \alpha ^{\prime }(t)+\beta ^{\prime }(t)+\delta
^{\prime }(t)\right) .
\end{eqnarray}

In all the calculations that follow, the equation of state parameters for
the isotropic and anisotropic fluids are set to be $\gamma =5$, $\gamma
_{1}=12$, $\gamma _{2}=18$, $\gamma _{3}=21$. These are representative
values that capture the essential behaviour that occurs whenever the
anisotropic ekpyrotic fluid is stiffer than the isotropic one ($\gamma
_{i}>\gamma $). Also, the results of the numerical integrations performed by
using different sets of initial conditions (Kasner and those that satisfy
the Friedmann constraint) have been plotted in the evolution of the scale
factors, $a(t)$, $b(t)$ and $c(t)$, using the definitions given in %
\Eref{expscale}.

As noted above, the Mixmaster behaviour seemed to occur in the form of
epochs \cite{Belinskii1970,Belinskii1972} with the memory of the `initial'
data being erased in successive Kasner epochs on approach to the
singularity. Thus, we choose Kasner-like `initial' conditions for the
variables we are integrating over. We then examine the effect of the
ultra-stiff fluids on their evolution. The initial values are as follows, 
\newline

$\fl\;\;\;\;\;\;\;\;\;\;%
\begin{array}{c|c|c|c|c|c|c|c}
\hline
x(\tau _{0}) & y(\tau _{0}) & H(\tau _{0}) & \alpha (\tau _{0}) & \beta
(\tau _{0}) & \delta (\tau _{0}) & \rho (\tau _{0}) & \mu (\tau _{0}) \\ 
\hline
m_{k1}-m_{k2} & m_{k1}-m_{k3} & \frac{(m_{k1}+m_{k2}+m_{k3})}{3} & m_{k1} & m_{k2} & 
m_{k3} & s & v \\ \hline
\end{array}%
$ $\vspace{20pt}$\linebreak where the three Kasner indices are expressed in
terms of the parameter $u$, as usual, by $m_{k1}=-u/(u^{2}+u+1)$, $%
m_{k2}=(u+1)/(u^{2}+u+1)$ and $m_{k3}=u(u+1)/(u^{2}+u+1)$ with $s=0.269943$, 
$v=0.20$. For the purposes of this computation, $u=-6\pi $ and $\tau
_{0}=-0.002$. The equations are evolved from $t=\tau _{0}$ to $t=\tau _{f}$,
where $\tau _{f}=-255$.\emph{\ }The singularity (if it occurs) is taken to
be at $t=0$ and indeed this is where all the quantities blow up in our
computation.\footnote{%
The integration is carried out in negative time as we are interested in a
contracting universe approaching the singularity and so we integrate
backwards in time. The sign of the time coordinate is not relevant as it can
be made positive by introducing a constant shift which would not affect our
results.} The values of the indices $m_{i}\hspace{3pt},\forall i=1,2,3,$
have been chosen according to the familiar Kasner vacuum parametrisation,
described for example in \cite{Belinskii1970}, and so satisfy $%
m_{1}+m_{2}+m_{3}=1=m_{1}^{2}+m_{2}^{2}+m_{3}^{2}$.

\begin{figure}[tbp]
\caption{Evolution of the scale factors for Bianchi type IX from Kasner-like
initial conditions towards the singularity. The universe contains isotropic (%
$\protect\rho >0$) ultra-stiff fluid only and no anisotropic pressure field
included,}
\label{kasner_scale_iso}\centering
\includegraphics[width=7cm,height=5cm]{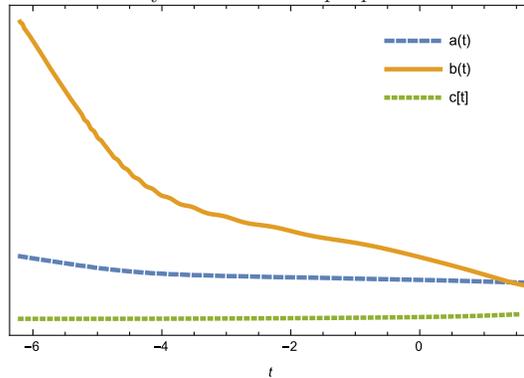}
\end{figure}

\begin{figure}[tbp]
\caption{Evolution of the scale factors for a Bianchi type IX universe from
Kasner-like initial conditions towards the singularity. The universe
contains an ultra-stiff isotropic fluid ($\protect\rho >0$) and a fluid with
anisotropic pressure.}
\label{kasner_scale_aniso}\centering
\includegraphics[width=7cm,height=5cm]{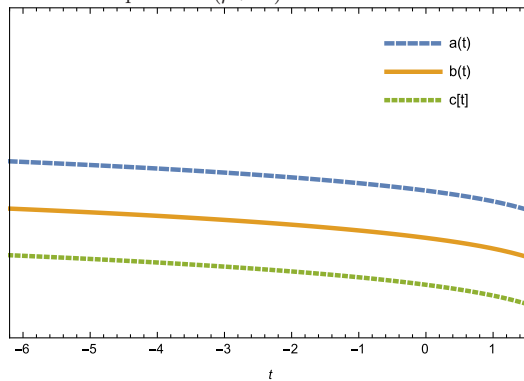}
\end{figure}

The initial hypersurface for the numerical computation has a geometry that
describes a flat, empty, anisotropic spacetime. This choice of initial
conditions do not exactly satisfy the Friedmann constraint exactly 
\Eref{friedmann_constraint_bianchi_ix}. The integration could only be
carried out for an ordinary (NEC-satisfying) ultra-stiff isotropic 
field. The numerical calculations show oscillations of the scale factors
before they are replaced by a nearly monotonic evolution towards the volume
minimum or singularity, as described in \cite{Belinskii1972}. This evolution
is depicted in \Fref{kasner_scale_iso} for the case including only the
isotropic field. For the case including both the isotropic ultra-stiff
field and an anisotropic pressure field (stiffer than the isotropic field)
the evolution is shown in \Fref{kasner_scale_aniso}.

On examining \Fref{kasner_scale_iso} and \Fref{kasner_scale_aniso}, we
observe the following features. The initial conditions used do not satisfy the 
Friedmann constraint and therefore later on in this section, the equations are solved again with initial conditions that do satisfy this constraint. There is more than one branch of solutions but at least one of the scale factors approaches the singularity in a slightly oscillatory manner as predicted in \cite{Belinskii1972}, while one of them stays nearly constant. On inclusion of the anisotropic energy density, the solutions seem to tend towards a contraction to a collapse, a fact that will be further verified in the case which takes into account the Friedmann constraint while picking initial conditions. In order to study the evolution of the shear and the near-singularity
behaviours of the scale factors and the Hubble rates, we try to find initial
conditions that satisfy the Friedmann constraint %
\eref{friedmann_constraint_bianchi_ix} and take into consideration the
curvature and the matter content of the spacetime. Accordingly, we choose,
\bigskip

{$%
\begin{array}{c|c|c|c|c|c|c|c}
\hline
x(\tau _{0}) & y(\tau _{0}) & H(\tau _{0}) & \alpha (\tau _{0}) & \beta
(\tau _{0}) & \delta (\tau _{0}) & \rho (\tau _{0}) & \mu (\tau _{0}) \\ 
\hline
m_{1}^{\prime }-m_{2}^{\prime } & m_{1}^{\prime }-m_{3}^{\prime } & 
(m_{1}^{\prime }+m_{2}^{\prime }+m_{3}^{\prime })/3 & m_{1} & m_{2} & m_{3}
& -s & v \\ \hline
\end{array}%
$ $\vspace{20pt}$\linebreak where $m_{1}^{\prime }=0.594778,m_{2}^{\prime
}=0.167825,m_{3}^{\prime
}=0.276172, m_{1}=1.19144,m_{2}=2.24155,m_{3}=1.22871,s=0.2175397$, and $v=0.20$; 
$\tau _{0}$ is the initial instant of time. For the present case, we have
chosen $\tau _{0}=1.6$. The equations are evolved from $t=\tau _{f}$ to $%
\tau _{0}=1.6$, where $\tau _{f}=-25$. These initial conditions satisfy
the Friedmann constraint \eref{friedmann_constraint_bianchi_ix} with an
error of only $\epsilon \sim O(10^{-8})$. }\emph{\ }

From the results of the numerical computation, we find the following
evolutionary features:

\subsubsection{ Scale-factor evolution}

\begin{figure}[tbp]
\caption{Scale factor evolution in Bianchi type IX with initial conditions
satisfying the Friedmann constraint with isotropic ultra-stiff ghost field ($%
\protect\rho <0$) and no ultra stiff anisotropic pressure field included}
\label{scale_factors_normal_iso}\centering
\includegraphics[width=7cm,height=5cm]{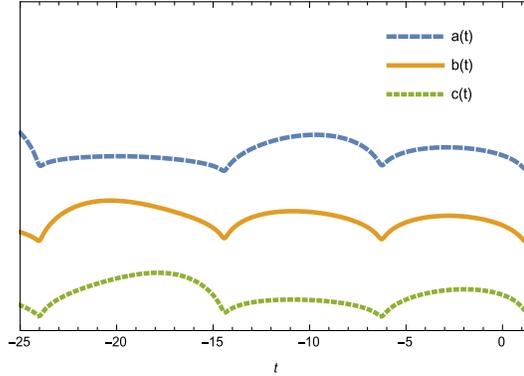}
\end{figure}

\begin{figure}[tbp]
\caption{Scale factor evolution in Bianchi type IX with initial conditions
satisfying the Friedmann constraint with isotropic ultra-stiff ghost field ($%
\protect\rho <0$) and anisotropic pressure ultra-stiff field included, on
approach to the singularity.}
\label{scale_factors_normal_aniso}\centering
\includegraphics[width=7cm,height=5cm]{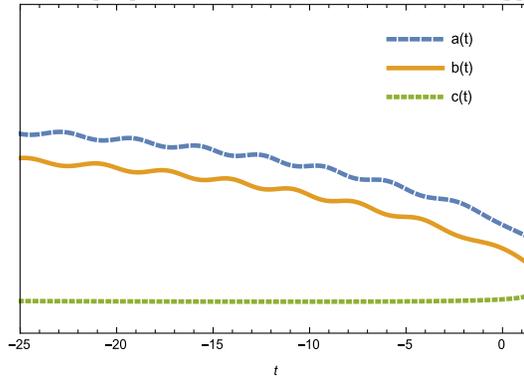}
\end{figure}

\begin{figure}[tbp]
\caption{Scale factor evolution in Bianchi type IX near the singularity with
initial conditions satisfying the Friedmann constraint with isotropic
ultra-stiff ghost field ($\protect\rho <0$) and no ultra-stiff anisotropic
pressure field included.}
\label{scale_factors_normal_iso_near}\centering
\includegraphics[width=7cm,height=5cm]{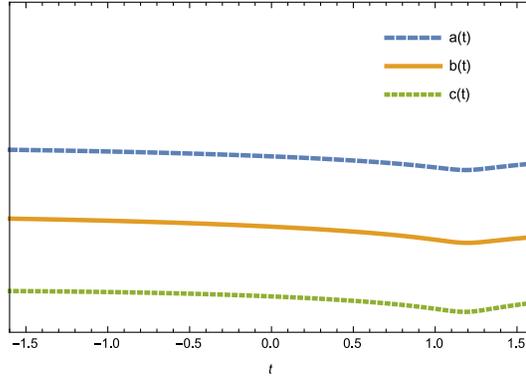}
\end{figure}

\begin{figure}[tbp]
\caption{Scale factor evolution in Bianchi type IX near the singularity with
initial conditions satisfying the Friedmann constraint with isotropic
ultra-stiff ghost field ($\protect\rho <0$) and ultra-stiff anisotropic
pressure field included. }
\label{scale_factors_normal_aniso_near}\centering
\includegraphics[width=7cm,height=5cm]{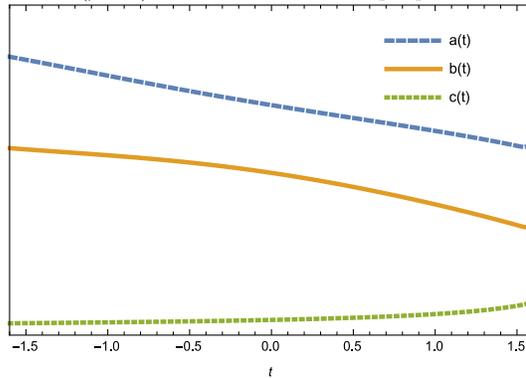}
\end{figure}

In the figures shown, that is in \Fref{scale_factors_normal_iso} and %
\Fref{scale_factors_normal_aniso}, the logarithms of the scale factors (i.e.,\
$\alpha ,\beta ,\delta $) have been plotted. The %
\Fref{scale_factors_normal_iso} shows the evolution of the scale factors
with the inclusion of only an isotropic ultra-stiff ghost field and %
\Fref{scale_factors_normal_aniso} shows the evolution of the scale factors
with the inclusion of both the isotropic, ultra-stiff ghost field and the
anisotropic pressure, which is also an ultra-stiff field and with greater
average stiffness than the ghost field. In the absence of the anisotropic pressure
field, we see that the scale factors undergo periodic bounces, with a phase of expansion, contraction
and a turnaround. On inclusion of the anisotropic pressure field, the periodic bouncing behaviour
is destroyed and the scale factor evolution seems to undergo gentle oscillations towards ultimately a collapse. One of the scale factors in the $c(t)$ direction remains almost constant throughout the evolution. We
study the near-singularity behaviour in more detail by focusing on the
evolution in a small time interval near to $t=0$. The %
\Fref{scale_factors_normal_iso_near} shows the evolution of the scale
factors with the inclusion of only an isotropic ultra-stiff ghost field very
close to the singularity and the \Fref{scale_factors_normal_aniso_near}
shows the evolution of the scale factors with the inclusion of both the
isotropic, ultra-stiff ghost field as well as the anisotropic pressure,
ultra-stiff (with greater stiffness than the ghost field) field very close
to the singularity.

%scale factors-constraint ic

%\clearpage

Near the singularity the solutions show the following behaviours. The scale
factors for the case including only the isotropic ghost fluid do not in fact
collapse to a singularity. They undergo a non-singular bounce, as expected
from our experience of the isotropic closed universe and that of the Kasner
universe with ghost field and radiation. However, the bounces in the three
directions seem to occur almost simultaneously. It is also interesting to note that if the stiffness of the
anisotropic fluid is increased, so that on the average it is stiffer than
the isotropic fluid, but is not stiff in one or two directions, then the
scale factors in those directions remain nearly constant. 
If the stiffness is less than that of the isotropic fluid,
or the initial conditions are such that the ultra-stiff anisotropic fluid is
negligible compared to the isotropic fluid density, they show similar
behaviour to the isotropic case and undergo a bounce after which the scale
factors all begin to re-expand. In all other cases, they contract until they
are very near the singularity. This means that near the expected
singularity, the isotropic fluid scale factors re-expand, but the scale
factors in the anisotropic fluid case all seem to contract towards a
singularity. In all cases, the shear for the case containing the isotropic
fluid alone is lower than when the anisotropic fluid is present.

%\clearpage

\subsubsection{Shear tensor}

\begin{figure}[tbp]
\caption{Evolution of the shear in a Bianchi type IX universe with initial
conditions satisfying the Friedmann constraint near the singularity.}
\label{shear_figure}\centering
\includegraphics[width=7cm,height=5cm]{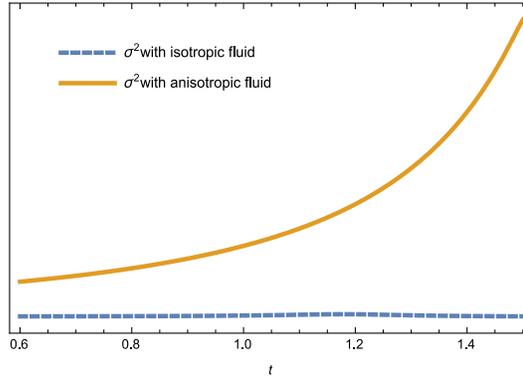}
\end{figure}

We had initially set out to investigate the effect of the anisotropic
pressure fluid on the evolution of the anisotropies. Thus, we next look at
the behaviour of the shear tensor on approach to the singularity (or
bounce). The shear tensor in the Bianchi type IX spacetime is given as, 
\begin{equation}
\sigma ^{2}=\frac{1}{6}\left\{ (H_{\alpha }-H_{\beta })^{2}+(H_{\beta
}-H_{\delta })^{2}+(H_{\delta }-H_{\alpha })^{2}\right\} ,
\end{equation}
where $H_{\alpha }=\dot{\alpha}$, $H_{\beta }=\dot{\beta}$ and $H_{\delta }=
\dot{\delta}$. We focus on the near-singularity behaviour of the shear
tensor. This evolution is shown in \Fref{shear_figure}. On examining the figure, we
find that with only the isotropic fluid present, the shear remains at a very small  and nearly constant positive value. However, when we include the anisotropic pressure fluid,
the shear rises and keeps rising to increasingly positive values until the
singularity is reached. This is true as long as the anisotropic pressure in
at least one direction is less stiff than the pressure of the isotropic
ghost fluid. This is equivalent to the requirement that one third of the
equation of state parameters $\gamma _{i}$ be less than the overall equation
of state parameter of the isotropic ultra-stiff ghost fluid. Thus, although
the anisotropic fluid may be ultra-stiff and stiffer than the isotropic
fluid, it may not be stiffer in a particular direction. This causes the
assumption that an energy source that behaves like ultra-stiff matter
suppresses the anisotropies near the singularity in a contracting universe
to break down. If the anisotropic pressures are stiffer than the isotropic
pressure in all three directions ($\gamma _{i}/3>\gamma $ for all $i=1,2,3$)
then the anisotropic stress is more greatly suppressed when the anisotropic
pressure fluid is included compared to when only the ghost isotropic fluid
is present. This is expected because it is simply the standard ekpyrotic
model with a stiffer fluid present in all three scale factor directions to
suppress the anisotropic stress. However, when the anisotropic pressure
fields have equations of state in each direction ensuring that the
anisotropic stress is not suppressed, then the universe fails to undergo a
bounce and re-expansion beyond the contracting phase. Instead, the
contraction accelerates towards a collapse singularity in the Weyl
curvature.

In addition to observing these trends, we also note the following general
features. When the stiffness of the anisotropic fluid is less than the
stiffness of the isotropic ghost fluid, the three scale factors all undergo
bounces. The stiffness of the anisotropic fluid determines when this bounce
occurs. If it becomes stiff on average ($\gamma _{\star }=2$), the scale
factors begin to oscillate on approach to the singularity. When the
anisotropic fluid is ultra-stiff on average (stiffer than the isotropic
ghost fluid), but its initial conditions are such that its density is
negligible or very small (less than half of the initial isotropic energy
density) the scale factors begin to show a turnaround at an expansion
minimum. The point of bounce is pushed towards the value at which the
turnaround occurs for the isotropic case, as the anisotropic energy density
is decreased.

\section{Conclusions}

The question of the growth of anisotropies in a contracting universe is a
challenge for cyclic theories of cosmology if they aim to replicate the
successes of the inflationary paradigm in explaining the present large-scale
isotropy of the universe. There are several types of anisotropy that need to
be investigated in order to ascertain the viability of cyclic cosmologies:
simple expansion rate anisotropy, spatial curvature anisotropy, and pressure
anisotropy. Simple expansion-rate anisotropies and 3-curvature anisotropies
can always be dominated by an ultra-stiff perfect fluid with equation of
state $p>\rho $. This is well appreciated and we confirm it here for the
Bianchi class A and type IX universes. In this paper, we have focussed on
the effects of pressure anisotropies in simple ekpyrotic \cite%
{Khoury2001,Khoury2002,Erickson2004} cyclic universe scenarios that are more
general and complicated than those first studied by Barrow and Yamamoto \cite%
{BYam}. Pressure anisotropies have been ignored in all other studies of
ekpyrotic and cyclic universes. It is important to include them in the
discussion because collisionless particles will be abundant near the Planck
scale where graviton production is rapid and asymptotically-free
interactions will not be in equilibrium.\emph{\ }In addition, we find that
if the average anisotropic pressure is allowed to exceed the energy density,
just as the isotropic pressure does in the ekpyrotic scenario, then an
isotropic singularity (or bounce) will be unstable unless the isotropic
density is overwhelmingly larger than the anisotropic density. The
anisotropic ultra-stiff fluid will drive a contracting universe to an
anisotropic singularity. Evolution from cycle to cycle will accumulate
anisotropic distortions to the dynamics.

More formally, we find that the anisotropy, even in the simplest case of a
Bianchi I universe with anisotropic pressures present, cannot be expressed
simply as a simple power-law evolution of the mean scale factor. Using a
phase-space analysis for the general field equations for the Bianchi Class A
group of cosmologies, we find that the presence of an ultra-stiff fluid with
anisotropic pressures prevents the isotropic Friedmann-Lema\^{\i}tre
universe from being an attractor for the initially contracting universe.
More specifically, we analysed the field equations in the case of the
Bianchi IX universe. We solved these equations numerically containing
ultra-stiff fluids with both anisotropic and isotropic (ghost) pressures.
The anisotropies grow when an anisotropic pressure fluid with dominant
stiffness is included: the universe contracts and hits a singularity. This
contrasts with the case containing only the isotropic ghost fluid, where the
universe undergoes a non-singular bounce. Our results confirm that the
inclusion of anisotropic pressures is essential in any general analysis of
cyclic cosmologies and their behaviour in the presence of deviations from
perfect expansion isotropy. They will be an important factor to consider in
all future iterations of the cyclic universe scenario in its several forms.

\textbf{Acknowledgements} JDB is supported by the STFC. CG is supported by
the Jawaharlal Nehru Memorial Trust Cambridge International Scholarship. The authors thank the referees for their very helpful comments, which have helped improve the current work considerably.

\appendix

\section{Orthonormal frame formalism and the Bianchi Class A models}

In this section we review the classification of the Bianchi cosmologies as given in \cite{WEllis}. The problem of classification of the Bianchi cosmologies can be seen as the problem of classifying the structure constants $C^{\mu}_{\alpha \beta}$ of the Lie algebra formed by the Killing vector fields(KVFs). These structure constants can be decomposed into a $2$-index symmetric object $\hat{n}_{\alpha \beta}$ and a $1$-index object $\hat{a}_{\alpha}$ as follows,
\begin{equation}
C^{\mu}_{\alpha \beta}=\epsilon _{\alpha \beta \nu}\hat{n}^{\mu \nu} +\hat{a}_{\alpha}\delta_{\beta}^{\mu}-\hat{a}_{\beta}\delta_{\alpha}^{\mu}
\end{equation}
such that $\hat{n}^{\mu \nu}=\hat{n}^{\nu \mu}$ and $\hat{a}^{\alpha}$ are constants. They also follow the identity,
\begin{equation}
\hat{n}^{\alpha \beta}\hat{a}_{\beta}=0
\end{equation}
The Lie algebras can thus be divided into Class A for $\hat{a}_{\alpha}=0$ and Class B for $\hat{a}_{\alpha} \neq 0$. 
In the standard language of the orthonormal frame formalism, we can define a unit timelike vector field $\textbf{u}$ and the projection tensor $h_{ab}$ which at each point projects into the space orthogonal to the unit timelike vector field. This projection tensor is given by,
\begin{equation}
h_{ab}=g_{ab}+u_au_b
\end{equation}
The covariant derivative of the timelike vector field can be divided into its irreducible parts,
\begin{equation}
u_{a;b}=\sigma _{ab}+\omega _{ab} +\frac{1}{3}\Theta h_{ab}-\dot{u}_a u_b
\end{equation}
where $\sigma _{ab}$ is the symmetric trace free, rate of shear tensor, $\omega _{ab}$ is the vorticity tensor and is antisymmetric, $\Theta$ is the rate of expansion scalar and $\dot{u}_a$ is the acceleration vector.
Relative to a group invariant orthonormal frame given by the unit normal to the group orbits and the basis vectors $\{\textbf{n},\textbf{e}_{\alpha}\}$, the EFEs are given above in equations \eref{efe1}, \eref{shear_1},  \eref{efe2} and \eref{friedman_constraint_1}.
The Jacobi identities are equation \eref{efe2} and the following equations,
\begin{eqnarray}
\dot{n}_{\alpha \beta }&=-Hn_{\alpha \beta }+2\sigma _{(\alpha }^{\mu
}n_{\beta )\mu }+2\epsilon _{(\alpha }^{\mu \nu }n_{\beta )\mu }\Omega _{\nu
}\\ 
\dot{a}_{\alpha}&=-Ha_{\alpha}-\sigma^{\beta}_{\alpha}a_{\beta}+\epsilon_{\alpha}^{\mu \nu}a_{\mu}\Omega_{\nu}\\ 
n_{\alpha}^{\beta}a_{\beta}&=0
\end{eqnarray}%
where $\Omega _{\nu}$ is the local angular velocity of the spatial frame with respect to the Fermi propagated spatial frame and can be expressed in terms of the components of the timelike vector field
$\textbf{u}$ as,
\begin{equation}
\Omega ^{\alpha}=\frac{1}{2} \epsilon ^{\alpha \mu \nu}e_{\mu}^ie_{\nu i;j}u^j
\end{equation} 
For our purposes, we have specialised to the case where the total stress energy tensor(isotropic stress energy tensor $+$ anisotropic part) is diagonal. Thus the more general case would include off diagonal elements of $P_{ab}$ as well and in this case $\Omega _{\nu}\neq0$. In our case however, we need only be concerned with universes where $\Omega _{\nu}=0$.

The spatial curvature terms can be defined as,
\begin{eqnarray}
^{3}S_{\alpha \beta}&=b_{\alpha \beta}-\frac{1}{3}(b^{\mu}_{\mu})\delta^{\alpha}_{\beta}-2\epsilon^{\mu \nu}_{(\alpha}n_{\beta)\mu}a_{\nu}\\ \nonumber
^{3}R&=-\frac{1}{2}b_{\mu}^{\mu}-6a_{\mu}a^{\mu} \nonumber
\end{eqnarray}
where,
\begin{equation}
b_{\alpha \beta}=2n_{\alpha}^{\mu}n_{\mu \beta}-(n_{\mu}^{\mu})n_{\alpha \beta}
\end{equation}
\section{New exact solution for $p=3\protect\rho $ fluid in Bianchi I
spacetime}

The field equations for the Bianchi I type spacetime are: 
\begin{eqnarray}
\ddot{\alpha}+\dot{\alpha}^{2}+\ddot{\beta}+\dot{\beta}^{2}+\dot{\alpha}\dot{%
\beta} &=&-p,  \label{a1} \\
\ddot{\beta}+\dot{\beta}^{2}+\ddot{\delta}+\dot{\delta}^{2}+\dot{\beta}\dot{%
\delta} &=&-p,  \label{a2} \\
\ddot{\delta}+\dot{\delta}^{2}+\ddot{\alpha}+\dot{\alpha}^{2}+\dot{\delta}%
\dot{\alpha} &=&-p,  \label{a3} \\
\dot{\alpha}\dot{\beta}+\dot{\beta}\dot{\delta}+\dot{\delta}\dot{\alpha}
&=&\rho ,  \label{a4}
\end{eqnarray}%
where the scale factors are expressed as $a(t)=\exp ({\alpha (t))}$, $%
b(t)=\exp ({\beta (t))}$, and $c(t)=\exp ({\delta (t))}$. Adding equations (%
\eref{a1})-(\eref{a3}), we get 
\begin{equation}
2(\ddot{\alpha}+\ddot{\beta}+\ddot{\delta})+2(\dot{\alpha}^{2}+\dot{\beta}%
^{2}+\dot{\delta}^{2})+(\dot{\alpha}\dot{\beta}+\dot{\beta}\dot{\delta}+\dot{%
\delta}\dot{\alpha})=-3p.
\end{equation}%
Using the formula $(a+b+c)^{2}=a^{2}+b^{2}+c^{2}+2(ab+bc+ca)$, we get, 
\begin{equation}
2(\ddot{\alpha}+\ddot{\beta}+\ddot{\delta})+2(\dot{\alpha}+\dot{\beta}+\dot{%
\delta})^{2}-4(\dot{\alpha}\dot{\beta}+\dot{\beta}\dot{\delta}+\dot{\delta}%
\dot{\alpha})+(\dot{\alpha}\dot{\beta}+\dot{\beta}\dot{\delta}+\dot{\delta}%
\dot{\alpha})=-3p.
\end{equation}%
Now substituting \Eref{a4} we get, 
\begin{equation}
2(\ddot{\alpha}+\ddot{\beta}+\ddot{\delta})+2(\dot{\alpha}+\dot{\beta}+\dot{%
\delta})^{2}-3\rho =-3p.  \label{equationref}
\end{equation}%
Defining the volume as $V\equiv \exp ({A)}$ where $A=(\alpha +\beta +\delta )
$ we get, 
\begin{equation}
\ddot{V}=3\rho _{0}V^{-3}.
\end{equation}%
Solving this gives 
\begin{equation}
V^{2}=C_{1}t^{2}+C_{2}t+C_{3}.
\end{equation}

Subtracting equations (\eref{a1}) and (\eref{a2}) we get, for
example, 
\begin{equation}
\ddot{\alpha}-\ddot{\beta}+3\dot{A}(\dot{\alpha}-\dot{\beta})=0,
\label{homogeneous}
\end{equation}%
and cyclic permutations. Thus we see that each of these combinations go as $%
V^{-1}$. We can write then, by integrating the above, 
\begin{equation}
\alpha -\beta =2l_{1}\mathrm{log}[\sqrt{t}+\sqrt{C_{2}+t}],
\end{equation}%
and 
\begin{equation}
\alpha -\delta =2l_{2}\mathrm{log}[\sqrt{t}+\sqrt{C_{2}+t}],
\end{equation}%
where $C_{1}=1,C_{3}=0$. We already know that 
\begin{equation}
(\alpha +\beta +\delta )=\frac{1}{2}\mathrm{log}[t^{2}+C_{2}t].
\end{equation}%
By using the fact that $3\alpha =(\alpha +\beta +\delta )+(\alpha -\beta
)+(\alpha -\delta )$, we obtain, 
\begin{equation}
3\alpha =\mathrm{log}\left[ (t^{2}+C_{2}t)^{1/2}(\sqrt{t}+\sqrt{%
C_{2}+t})^{2(l_{1}+l_{2})}\right] .
\end{equation}%
Thus, 
\begin{equation}
a(t)=\left( (t^{2}+C_{2}t)^{1/2}(\sqrt{t}+\sqrt{C_{2}+t}%
)^{2(l_{1}+l_{2})}\right) ^{1/3},
\end{equation}%
\begin{equation}
b(t)=\left( (t^{2}+C_{2}t)^{1/2}(\sqrt{t}+\sqrt{C_{2}+t}%
)^{2(l_{2}-2l_{1})}\right) ^{1/3},
\end{equation}%
\begin{equation}
c(t)=\left( (t^{2}+C_{2}t)^{1/2}(\sqrt{t}+\sqrt{C_{2}+t}%
)^{2(l_{1}-2l_{2})}\right) ^{1/3}.
\end{equation}

From the Friedmann constraint equation at late times (where $\rho
\rightarrow 0$), we get the following constraint, 
\begin{equation}\label{k_cons_1}
l_{1}^{2}+l_{2}^{2}-l_{1}l_{2}=1.
\end{equation}

We label the indices in the solutions for the scale factors as follows, 
\begin{eqnarray}
3q_{1} &=&1+l_{1}+l_{2}, \label{k_cons_2}\\
3q_{2} &=&1+l_{2}-2l_{1},\label{k_cons_3} \\
3q_{3} &=&1+l_{1}-2l_{2}.\label{k_cons_4}
\end{eqnarray}%
Therefore, we have the full solution:

\begin{equation}
a(t)=\left( (t^{2}+C_{2}t)^{1/2}(\sqrt{t}+\sqrt{C_{2}+t}%
)^{2(3q_{1}-1)}\right) ^{1/3},
\end{equation}%
\begin{equation}
b(t)=\left( (t^{2}+C_{2}t)^{1/2}(\sqrt{t}+\sqrt{C_{2}+t}%
)^{2(3q_{2}-1)}\right) ^{1/3},
\end{equation}%
\begin{equation}
c(t)=\left( (t^{2}+C_{2}t)^{1/2}(\sqrt{t}+\sqrt{C_{2}+t}%
)^{2(3q_{3}-1)}\right) ^{1/3}.
\end{equation}

We see that at early times this solution tends to the flat Friedmann
solution for $p=3\rho $ fluid ($a\sim t^{1/6}$, $b\sim t^{1/6}$, $c\sim
t^{1/6}$) as $t\rightarrow 0$, and at late times approaches the vacuum
Kasner solution $a\sim t^{q_{1}}$, $b\sim t^{q_{2}}$ and $c\sim t^{q_{3}},$%
with $\sum_{i}q_{i}=1=\sum_{i}q_{i}^{2}$, as $t\rightarrow \infty .$ These facts
can be seen to be true from equations \eref{k_cons_2} to \eref{k_cons_4} and from equation\eref{k_cons_1}
respectively. Thus,
this solution provides a simple exact description of the transition from an
isotropic initial state to a Kasner-like anisotropic future in a particular
case. It displays the opposite evolutionary trend to the evolution of a $%
0\leq p<\rho $ perfect-fluid model.

\section{Fixed points}

In order to perform the stability analysis on the Bianchi Class A system, we
need to identify the fixed points of the system. They have been presented in
a tabular form in \Tref{fixed_points}. The explicit forms of the relevant fixed points are given below.
\begin{eqnarray*}
\fl\Sigma _{\!_{P_{II}}}^{\!^{(1)}} =\frac{1}{8}(3\gamma -2)\\
\fl n_{P_{II}}^{(1)} =\frac{3}{4}[(2-\gamma )(3\gamma -2)]^{1/2}, \\
\fl\Omega _{P_{II}}^{(1)} =\frac{3}{16}(\gamma -6), \\
\fl \Sigma _{PVI}^{(1)}=-\frac{1}{4}(3\gamma -2)\\
\fl n_{PVI}^{(1)}=\pm \frac{3}{4}\sqrt{(2-\gamma)(3\gamma -2)}\\
\fl \Omega _{PVI}^{(1)}=-\frac{3}{4}(2-\gamma)\\t
\fl\Sigma _{\!_{K-}}^{\!^{2}}+\Sigma _{\!_{K+}}^{\!^{2}} =1 \\
\fl\Sigma_{\mathcal{A}_1}^2=\frac{4(\mathcal{P}_+^2 +\mathcal{P}_- ^2)}{(\gamma_{\star}-2)^2}\\
\fl \Sigma _+ ^{\mathcal{A}_2}=\frac{1}{2}\frac{\mathcal{P}_+(\gamma -\gamma _{\star})}{\mathcal{P}_+^2+\mathcal{P}_-^2}\\
\fl \Sigma _- ^{\mathcal{A}_2}=\frac{1}{2}\frac{\mathcal{P}_-(\gamma -\gamma _{\star})}{\mathcal{P}_+^2+\mathcal{P}_-^2}\\
\fl\Omega^{\mathcal{A}_2}=-1+\frac{1}{4}\frac{(2-\gamma _{\star})(\gamma -\gamma _{\star})}{\mathcal{P}_+^2+\mathcal{P}_-^2}\\
\fl Z^{\mathcal{A}_2}=\frac{1}{4}\frac{(2-\gamma)(\gamma -\gamma _{\star})}{\mathcal{P}_+^2+\mathcal{P}_-^2}\\
\end{eqnarray*}%
On examination of the forms of the fixed points, we find that only the FL,
Kasner and the $\mathcal{A}_{1}$ and $\mathcal{A}_{2}$ points are physical for
the case considered, that is, for ultra stiff fields, with $\gamma >2$.

\section{Equations for the Bianchi IX numerical computation}

In \Sref{sec:bianchi_ix}, a new system of variables was introduced to make
the numerical computation of the system of Einstein's equations simpler by
reducing them to first-order differential equations. They are written
explicitly as follows. In all of the following $\{x,y,H,\alpha,\beta,\delta,\rho,\mu\}$ are functions of time $t$.

\begin{equation}
\fl x^{\prime }+3Hx=\left( \gamma _{1}-\gamma
_{2}\right) \mu+ e^{-2\beta}-e^{-2\alpha}+e^{-2\delta}\left(e^{2(\beta-\alpha)}-e^{2(\alpha -\beta)}\right)
\end{equation}
\begin{equation}
\fl y'+3Hy=(\gamma _1-\gamma _3)\mu -e^{-2\alpha}+e^{-2\delta}+e^{-2\beta}\left(e^{2(\delta -\alpha)}-e^{2(\alpha -\delta)}\right)
\end{equation}
\begin{eqnarray}
\fl 6(H'+3H^2)=3\left(2-\gamma\right)\rho+3\left(2-\gamma_{\star}\right)\mu-2 e^{-2\alpha}-2 e^{-2\beta}-2 e^{-2\delta}\\ \nonumber
+e^{2(\alpha-\beta-\delta)}+e^{2(\beta-\alpha-\delta)}+e^{2(\delta-\alpha-\beta)}
\end{eqnarray}
\begin{equation}
\fl\rho ^{\prime }=-3\gamma H\rho , \\
\nonumber
\end{equation}%
\begin{eqnarray}
\fl\mu ' =-3\mu H-\frac{(\gamma _{1}-1)}{3}(x+y+3H)%
\mu-\frac{(\gamma _{2}-1)}{3}(3H-2x+y)\mu\\ \nonumber
-\frac{(\gamma _{3}-1)}{3}(x-2y+3H)%
\mu
\end{eqnarray}%
\begin{equation}
\fl\alpha ^{\prime }=\frac{1}{3}(3H+x+y), \\
\nonumber
\end{equation}%
\begin{equation}
\fl\beta ^{\prime }=\frac{1}{3}(3H-2x+y), \\
\nonumber
\end{equation}%
\begin{equation}
\fl\delta ^{\prime }=\frac{1}{3}(3H+x-2y). \\
\nonumber
\end{equation}
\pagebreak
%\bibliographystyle{unsrtnat}
%\bibliography{Bianchi_Class_A_and_type_IX}

\end{document}